\documentclass[sn-basic]{sn-jnl}

\usepackage{amssymb,amsmath}
\usepackage{hyperref}
\usepackage[misc]{ifsym}
\usepackage{natbib}
\usepackage{subfigure}
\usepackage{times}

\theoremstyle{thmstyleone}%
\theoremstyle{thmstyletwo}%

\theoremstyle{thmstylethree}%

\begin{document}

\title{How many preprints have actually been printed and why: a case study of computer science preprints on arXiv}


\author[1]{\fnm{Jialiang} \sur{Lin}}\email{me@linjialiang.net}

\author[1]{\fnm{Yao} \sur{Yu}}\email{eloria1995@gmail.com}

\author[1]{\fnm{Yu} \sur{Zhou}}\email{xmuzy@stu.xmu.edu.cn}

\author[1]{\fnm{Zhiyang} \sur{Zhou}}\email{zyzhou95@gmail.com}

\author*[1]{\fnm{Xiaodong} \sur{Shi}}\email{mandel@xmu.edu.cn}

\affil[1]{\orgdiv{School of Informatics}, \orgname{Xiamen University}, \orgaddress{\city{Xiamen}, \country{China}}}


\abstract{Preprints play an increasingly critical role in academic communities. There are many reasons driving researchers to post their manuscripts to preprint servers before formal submission to journals or conferences, but the use of preprints has also sparked considerable controversy, especially surrounding the claim of priority. In this paper, a case study of computer science preprints submitted to arXiv from 2008 to 2017 is conducted to quantify how many preprints have eventually been printed in peer-reviewed venues. Among those published manuscripts, some are published under different titles and without an update to their preprints on arXiv. In the case of these manuscripts, the traditional fuzzy matching method is incapable of mapping the preprint to the final published version. In view of this issue, we introduce a semantics-based mapping method with the employment of Bidirectional Encoder Representations from Transformers (BERT). With this new mapping method and a plurality of data sources, we find that 66\% of all sampled preprints are published under unchanged titles and 11\% are published under different titles and with other modifications. A further analysis was then performed to investigate why these preprints but not others were accepted for publication. Our comparison reveals that in the field of computer science, published preprints feature adequate revisions, multiple authorship, detailed abstract and introduction, extensive and authoritative references and available source code.}

\keywords{Preprint, ArXiv, Digital repositories}



\maketitle

\section{Introduction}

The traditional pipeline for academic publication is highly time-consuming~\citep{bjork-publishing-2013}. The whole publishing process, from doing research, writing a paper, submitting for peer review, revising or rewriting if rejected all the way to final publishing, can be a weary march that costs several or even a dozen months. Because of this, some researchers who cannot afford to wait then turn to conferences for publication since the process can be reduced to only a few months. But still there are lots of other researchers out there who are eager to share their research results as soon as possible. To them, even a few months would be too long. There thus came into existence those popular preprint servers. ``A preprint is a complete scientific manuscript (often one also being submitted to a peer-reviewed journal) that is uploaded by the authors to a public server without formal review''~\citep{berg-preprints-2016}. Users of preprint servers can post their manuscripts without rigorous peer review but only with a brief censoring. Even though ``preprint'' indicates the concept of pre-submission before publication, there also exist a large number of post-prints that are submitted to preprint servers after publication. Therefore in this paper, ``preprint'' is defined as ``e-print''~\citep{kling-internet-2004,brody-earlier-2006} that implies both ``pre-print'' and ``post-print'' on preprint servers. The word ``unpublished'' is used to describe the state of a preprint in which it has not yet been accepted for any type of publications; the words ``published'' and ``printed'' are used to describe the state of a preprint in which it has been peer-reviewed and formally published in journal, conference, book, report or other types of publications.

ArXiv\footnote{\url{https://arxiv.org/}.}~\citep{ginsparg-arxiv-2011}, founded in 1991, is a preprint server in the field of science and engineering. From its inception to 2014, arXiv housed a total of 1 million manuscripts after 23 years of development~\citep{van-arxiv-2014} and in 2019, it received an average of about 13,000 monthly submissions.\footnote{\url{https://arxiv.org/stats/monthly_submissions}.} Computing Research Repository (CoRR)~\citep{halpern-corr-2000} is a respected component of arXiv. This repository covers various categories of computer science (CS) and enjoys a rapid increase in submissions. CoRR now functions as the most important preprint server in the CS field.

In addition to arXiv, there are a considerable number of other preprint servers for different fields. BioRxiv\footnote{\url{https://www.biorxiv.org/}.} is a platform for unpublished preprints, especially those in life science. Unlike arXiv, bioRxiv assigns DOI to its preprints for citation and cooperates with journals in a way that enables authors to submit their manuscripts directly to a journal's submission system through it. Social Science Research Network (SSRN)\footnote{\url{https://www.ssrn.com/}.} is a repository originally developed for social science and humanities. It was later extended to cover other fields in science and engineering, such as biology, chemistry, and computer science. This platform allows users to upload their unpublished preprints directly and it also accepts published papers. Humanities Commons\footnote{\url{https://hcommons.org/}.} is a platform created by the Modern Language Association for the field of humanities. It serves as a network for humanities scholars to post new publications and disseminate research results. Preprints\footnote{\url{https://www.preprints.org/}.} is a multidisciplinary preprint server supported by the open access publisher MDPI\footnote{MDPI is an acronym referring to two related organizations, Molecular Diversity Preservation International and Multidisciplinary Digital Publishing Institute.} that provides immediate accessibility to scientific manuscripts in all fields of research. This server provides users with the number of views and downloads that a preprint has received. Users can also make comments on the preprints.

Such prosperous preprint servers are driven by several forces because submitting manuscripts to them comes with many benefits. Firstly, preprints give a record of priority. In many cases of research, researchers might conduct studies under similar topics and methods, but unfortunately this similarity can lead to fierce controversy on priority just like what happened between Isaac Newton and Gottfried Wilhelm Leibniz over the nearly simultaneous invention of calculus. Therefore, it is vital for researchers to publish their original ideas and research results in time in the academic field. Secondly, credit is given to preprints for enabling more feedback. Good research and papers imply rounds of refinement. Within traditional peer review, authors can only get limited rounds of feedback from only a handful of reviewers and editors. However, publishing the manuscripts in early stage can elicit discussion and feedback from the whole community. Thirdly, a preprint can function as an attention grabber. Most preprint servers provide daily notification service to service-subscribers, sending them lists of latest submissions and updated manuscripts. There are studies~\citep{davis-does-2007,feldman-citation-2018} revealing that published papers with preprints submitted before publication gain more citations than those without. Some researchers, in order to present their work to more people, choose to make submissions to preprint servers for public access even after their work has been accepted by peer review.

However, preprint servers also lead to widespread controversy~\citep{vale-accelerating-2015,annesley-biomedical-2017}. For one thing, there is no guarantee on the quality of the non-peer-reviewed preprints. Even though some preprint servers, such as arXiv and bioRxiv do perform inspection on the submissions, this inspection only targets non-scientific content, plagiarism, and fouled words, and cannot ensure the internal academic quality. Unfinished or even fraudulent preprints, which might be submitted just for scooping, cast a detrimental influence on refereed publication. Whether these kinds of preprints can be considered as a claim of priority or not remains open to question. For another, a simplified process might lead to a growing burst in the number of academic submissions, which could then become a burden for researchers to distinguish between good and bad.

There exists such an enormous number of preprints and yet it remains unclear how many preprints have actually been printed and why. This paper sets out to answer these two questions by conducting a case study on CS related preprints on arXiv from 2008 to 2017. Our main contributions lie in:

\begin{enumerate}
	\item A BERT-based method and a related dataset are introduced to map preprints to their published versions under different titles and with other modifications. Our method achieves an improvement of 56\% in accuracy over the compared method.
	\item Mapping was conducted from 141,961 sampled preprints to their published versions one by one. Statistical analyses are performed on different aspects including published type, subject category, publication venue, submission stage, and citation count.
	\item Common features of published preprints are identified by in-depth comparisons conducted between the published and the unpublished. Practical suggestions for future academic writing are provided based on the findings and analysis.
\end{enumerate}

\section{Related work}

Former studies on preprints basically cover the areas of citation, publications, impact, preprint servers, and peer review publication.

ArXiv provides its users with usage statistics,\footnote{\url{https://arxiv.org/help/stats}.} but the information is limited to the statistics of submission, access and download. The research of \citet{davis-does-2007} is an early work that analyzed the correlation between the submission of a preprint and the citation and official download counts of its final publication in the field of mathematics. Their study identified 511 (18.5\%) published preprints out of 2,765 sampled journal papers on arXiv. However, the authors did not mention which method they employed to map the published papers to their preprints on arXiv.

\citet{lariviere-arxiv-2014} analyzed arXiv preprints in all subjects (computer science, mathematics, physics, etc.) and the corresponding published versions on Web of Science (WoS). However, most of the conference papers in CS were excluded from their study since they were not indexed in WoS. In the domain of CS, conference papers, especially those submitted to top conferences, play a more vital role than journal papers~\citep{vrettas-conferences-2015}. Such exclusion implies incompleteness of study in the world of CS.

On the contrary, \citet{sutton-popularity-2017} analyzed papers published in top CS conferences and found that in the year of 2017, 23\% of these conference papers have been submitted to arXiv. The study also shows that 56\% of the above arXiv-deposited papers were submitted before or during the review process. Though with interesting findings, this study only deals with papers of top conferences. In the research, they checked off the published papers listed in the conference proceedings one by one to identify whether these papers were submitted to arXiv before or not. Such a method leaves out those arXiv-deposited papers that are published in other academic venues. Although the CS community generally attaches greater importance to top conferences in this field, journals and other types of publications remain an indispensable and significant part that should not be ignored.

\citet{feldman-citation-2018} explored whether arXiv-deposited papers can gain more citations in the field of top CS conferences. In their research, they adopted a mapping method similar to that of \citet{sutton-popularity-2017} to map the preprints on arXiv to their corresponding accepted papers by matching paper titles on the conference paper lists with their metadata on arXiv.

To our best knowledge, there exists no preprint study that covers all basic types of academic publications, including conference, journal, book chapter, and others in the field of CS. Moreover, using correspondence matching or traditional fuzzy mapping to match preprints and their publications may lead to impreciseness. This paper would like to fill this gap by introducing a BERT-based matching method that can capture the semantic information of titles. Unlike most previous studies that only checked off a limited list of publishing papers against arXiv records, we embark on a different route as we check off the preprints on arXiv against other databases for matching.

\section{Data sources}
In this section, we describe the data sources used in our research. This research samples preprints that are first submitted to arXiv within the time period from 2008 to 2017 and fall under at least one category that starts with the category prefix ``cs.'' (an indicator for the field of computer science). A total of 141,961 preprints are thus identified according to these two criteria. The reason why we do not set the time range of sample data till the time of writing this paper is that it may take a long time for a preprint to be reviewed, revised, and published. We need to leave enough turnaround time for formal publication. Multiple data sources of arXiv, Crossref, DBLP, Google Scholar, and Papers With Code are used to support our research.

\subsection{ArXiv}
Apart from web page access, arXiv also opens its metadata to public access via Application Programming Interface (API).\footnote{\url{https://arxiv.org/help/api}.} This metadata includes article ID, version number, title, authors, categories, abstract, created date, and updated date. Some preprints also provide optional data like Digital Object Identifier (DOI), journal references, comments, etc. If a preprint has been updated before, version history will also be presented. Users can use Amazon S3 for bulk download of packed PDF files.\footnote{\url{https://arxiv.org/help/bulk_data}.} We harvested the sample data of both metadata and PDF files from arXiv in July 2019.

\subsection{Crossref}
Crossref,\footnote{\url{https://www.crossref.org/}.} launched in 1999, aims to establish cross-publisher citation linking for academic publications~\citep{lammey-crossref-2014}. As an official DOI registration agency of the International DOI Foundation, Crossref allows linking among a vast number of publications of different content types including journals, conference proceedings, books, data sets, etc. It works with thousands of publishers to provide authorized access to their metadata including DOI, publication date, and other basic information. Via APIs\footnote{\url{https://github.com/CrossRef/rest-api-doc}.} its metadata is also made available for free public access with publication titles or DOIs. We invoked Crossref APIs and stored the related data in August 2019.

\subsection{DBLP}
The Digital Bibliography and Library Project (DBLP) with the new title of ``The DBLP Computer Science Bibliography''\footnote{\url{https://dblp.org/}.}~\citep{ley-dblp-2002}, is a famous bibliography website centering around CS. It has been proved that its database indexes the largest amount of CS papers~\citep{cavacini-what-2015}. This website only stores basic publication information without abstract. In addition to publication in peer-reviewed venues, DBLP also indexes preprints in CoRR. Its data is provided in an Extensible Markup Language (XML) file.\footnote{\url{https://dblp.org/xml/}.} Our research is based on the version of ``dblp-2019-10-01''.

\subsection{Google Scholar}
Thanks to the powerful search and analysis technologies of Google, Google Scholar\footnote{\url{https://scholar.google.com/}.} plays a leading role in academic literature analysis and retrieval platform service. Unlike other citation analysis platforms that only provide indexes to journal papers with high impact factors and usually written in English, Google Scholar indexes a wide range of academic documents (journal papers, conference papers, books, theses, etc.) written in various languages~\citep{kousha-sources-2008,martin-google-2018}. Google Scholar citations can thus be used to fully reflect the overall citation of a paper~\citep{martin-can-2017}. Since APIs are not available on Google Scholar, these data were crawled in the 2 months of August and September in 2019.

\subsection{Papers With Code}
Papers With Code\footnote{\url{https://paperswithcode.com/}.} links the source code to arXiv papers. On one hand, it labels data automatically with the use of Natural Language Processing technology to analyze paper contents and extract evaluation metrics. On the other hand, it also labels data by hand. The website provides daily-updated metadata in the format of JavaScript Object Notation (JSON). We downloaded the file on October 14th, 2019.

Since the process of crawling, downloading, and processing data is intensely time-consuming, our data collection process lasted several months. Nevertheless, the dates of our sampled data were generated at least 18 months ago, and thus these data are subject to minor variation only. Therefore, the slightly prolonged duration of data collection had little effect on our research results.

\section{Methods}
\label{sec:methods}

Presented in this section are analysis methods on identifying the number of sampled preprints that are accepted for publication in peer-reviewed venues. There are three cases of published preprints: (1) preprints published under the same titles with DOIs or names of the specific published venues provided by arXiv; (2) preprints published under the same titles without DOIs or published venues provided; (3) preprints published with their titles changed without DOIs or published venues provided.

\subsection{Case one}
ArXiv cooperates with Inspire (formerly SPIRES) to provide automatic update to DOI information and journal references if a preprint is published.\footnote{\url{https://arxiv.org/help/bib_feed}.} In addition, it also encourages authors to update this information themselves for their accepted manuscripts.\footnote{\url{https://arxiv.org/help/jref}.} A total of 28.7\% of our sampled data were confirmed to be published in peer-reviewed venues with 22.1\% offering DOI information in their metadata and 6.6\% with specific publication venues but no DOI.

\subsection{Case Two}
For preprints not included in the first case, we first conducted a search on Crossref or DBLP to examine how many sampled preprints are published under the same titles with the original first author appearing in the authorship. If a search result is in accord with such conditions, it is considered the published version of its preprint on arXiv. Through this process, 37.0\% of the sampled preprints are identified.

\subsection{Case Three}
The remaining sampled preprints were taken into consideration according to the following three different situations:

\begin{enumerate}
	\item Preprints that are not submitted to or accepted by a peer-reviewed venue.
	\item Preprints that are published in some venues that are not indexed by Crossref or DBLP.
	\item Preprints that are published with changed titles and content after peer review but without timely version update on arXiv.
\end{enumerate}

In the case of Situation 3, these preprints, with revisions on the title, content, and even authors, are hard to be identified with simple string matching. In view of this issue, we designed a classification model and constructed a special dataset to conduct pair matching between the sampled preprints and their revised published versions.

\subsubsection{Dataset for the classification model}
\label{sec:dataset}

A special dataset is constructed for our binary classification model. Both our positive and negative samples are composed of the following three fields: title pairs (preprint, candidate), author pairs (preprint, candidate), and a True or False label that indicates whether the candidate is the modified version of the preprint.\footnote{We originally added abstract pairs (preprint, candidate) to our dataset, but we found that only a little part of data of Crossref had abstracts when the model had actually been used, so the abstract pairs were removed in our model.}

To avoid overlap among training, development, and test data, the training set only includes data which are not included in the development or test set. To be more specific, they are preprints under CS category within the time period from 1991 through 2007, from 2018 through July 2019, and preprints from 2008 through 2017 but fall under no category in CS.

The arXiv API presents version history information of each preprint by attaching a version number as a suffix to the file names, like v1, v2. These data directly indicate the version sequence, and were thus gathered to create positive samples.

For each preprint, there may be several committed versions with title changed or not. We choose every two of different titles for each preprint to form a positive sample pair, which means these two titles are different but belong to the same preprint. An example is given as follows. The sampled preprint\footnote{\url{https://arxiv.org/abs/1901.07213}.} changes titles for each of the four submissions:

\begin{itemize}
	\item v1: Fully Convolutional Network-based Multi-Task Learning for Rectum and Rectal Cancer Segmentation
	\item v2: Multi-Task Learning with a Fully Convolutional Network for Rectum and Rectal Cancer Segmentation
	\item v3: A Fully Convolutional Network for Rectal Cancer Segmentation
	\item v4: Reducing the Model Variance of Rectal Cancer Segmentation Network
\end{itemize}

Titles of the four versions are different from each other, and altogether we can draw six positive sample pairs from them. The corresponding authors of each version were also extracted to compose input pairs.

To compose negative samples, each preprint title was submitted as a search query to Crossref for the first ten results returned, with information on title and author. Among the results, if one or more authors of a result are matched with the query's, this result will be removed. For the conditions of unmatched, they will be paired with the query one by one. The corresponding authors of the papers in query results were also gathered to form input pairs.

There are a total of 40k samples in the training set, 5k in the development set, and 5k in the test set.

\subsubsection{Classification model}
Bidirectional Encoder Representations from Transformers~\citep{devlin-bert-2019} is a state-of-the-art framework for word encoding. Its architecture enables BERT to learn contextual word embedding from both left to right and right to left. Our model is constructed on the BERT-based SciBERT~\citep{beltagy-scibert-2019}. It is a domain-specific model for scientific papers developed through fine-tuned BERT. Unlike BERT, SciBERT is trained on 1.14 million papers in the domain of CS and biomedical exclusively. Experiments show that SciBERT can achieve better performance than BERT on scientific text.

\begin{figure}[h]
	\centering
	\includegraphics[scale=0.78]{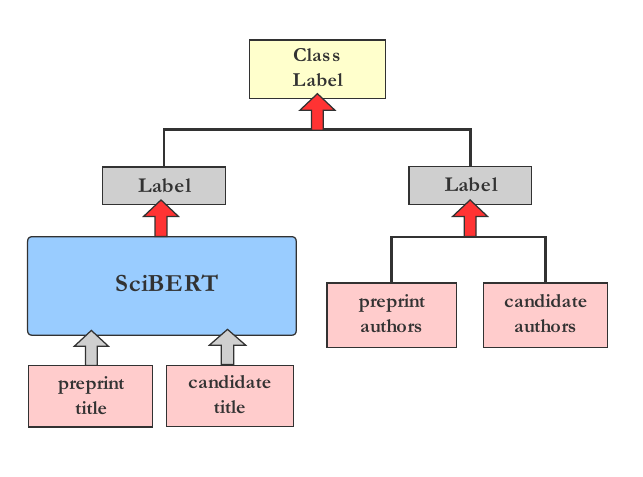}
	\caption{Structure of the proposed classification model}
	\label{fig:similar-model}
\end{figure}

SciBERT is fed with a pair of titles from two different papers (a preprint and its candidate to be checked) and outputs a probability value that indicates the similarity between these two papers. If the output value is higher than 0.5, the label of this pair will be set to True, otherwise to False. For the author pair from the above two papers, if the first author of the preprint matches one of the authors of the candidate, the label of the author pair will be set to True, otherwise to False. An ``and'' operation is then performed on these two boolean values to output a final label that indicates the likely relation between these two papers. See Fig.~\ref{fig:similar-model} for the detailed structure.

\subsubsection{Model results}
The classification model we propose yields an accuracy of 0.78 and an overall F1-score of 0.72. The work of \citet{lariviere-arxiv-2014} was chosen as the compared method which maps preprints and their published versions with different means of fuzzy matching. Our method outperforms the compared method in both accuracy and F1-score on the test set mentioned in Section~\ref{sec:dataset}. See Table~\ref{table:performance-comparison} for detailed information. With our model, we can identify preprints published under changed titles better. Finally, 11.4\% preprints of sampled data are mapped with their published version under changed title.

\begin{table}[h]
	\caption{Comparisons of accuracy and F1-score with the compared method}
	\label{table:performance-comparison}
	\centering
	\begin{tabular}{llll}
	\hline\noalign{\smallskip}
					    	   		&  Accuracy   			&  F1-score      		 \\
    \noalign{\smallskip}\hline\noalign{\smallskip}
		Proposed method    			        &  0.78           &  0.72         		 \\
		Compared method~\citep{lariviere-arxiv-2014}  &  0.50           &  0.67         		 \\
	\noalign{\smallskip}\hline
	\end{tabular}
\end{table}

\section{Statistics and analysis}
\label{sec:statistics-analysis}

\subsection{Published type}
\begin{figure}[h]
	\centering
	\includegraphics[scale=0.5]{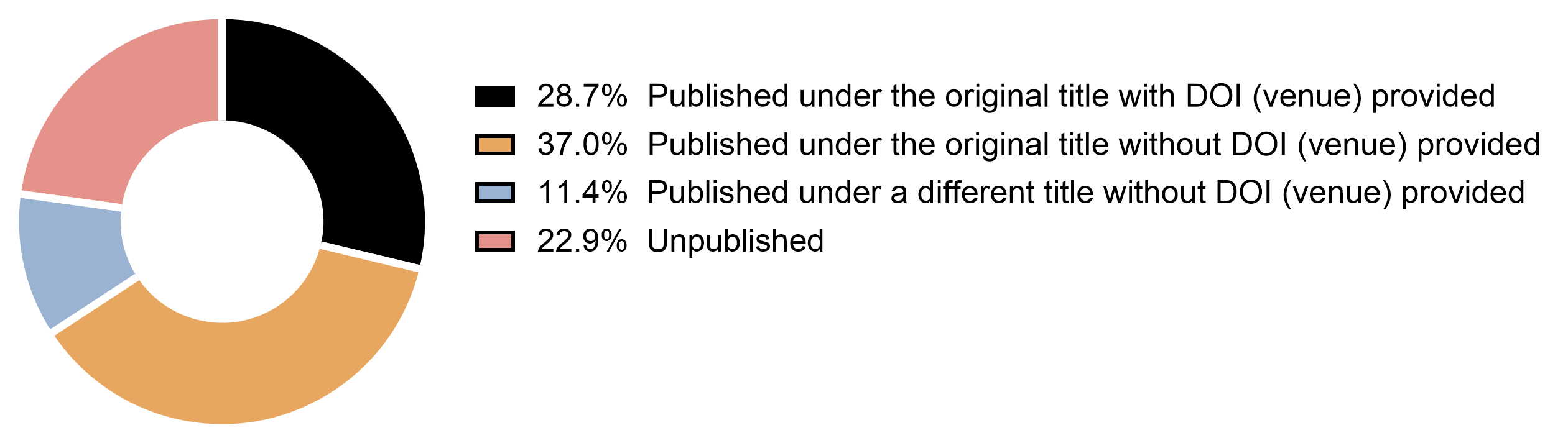}
	\caption{Distribution of preprints by published type}
	\label{fig:stat-by-published-type}
\end{figure}

According to the above data, 65.7\% of CS related preprints submitted to arXiv within the time period from 2008 through 2017 have been published in peer-reviewed venues with the same titles, and 11.4\% are published under changed titles and with other modifications. The whole sampled data are categorized into four types. See Fig.~\ref{fig:stat-by-published-type} for detailed information.

We estimated that nearly a quarter of the sampled preprints on arXiv have not been published and we performed an analysis to figure out the reasons behind, which are listed as follows:

Firstly, some unpublished preprints are strongly related to arXiv itself~\citep{warner-open-2001,rieger-arxiv-2016}. These preprints are written by the founders or administrators of arXiv to introduce its history, status quo, and development. From the very beginning, these authors only have the intention to present these preprints on arXiv. Secondly, there are also preprints that have indeed been submitted for peer review but fail to be accepted for publication.\footnote{Preprints which fall under this condition have statements like ``submitted to (a certain journal)'' or ``submitted to (a certain conference)'' included in their metadata. However, up until this paper is written, no corresponding records can be found in any journal or conference proceedings. The results are further confirmed with the method presented in Section~\ref{sec:methods}. Therefore, we reach the conclusion that these preprints submitted fail to be accepted.}

\begin{figure}[h]
	\centering
	\includegraphics[scale=0.54]{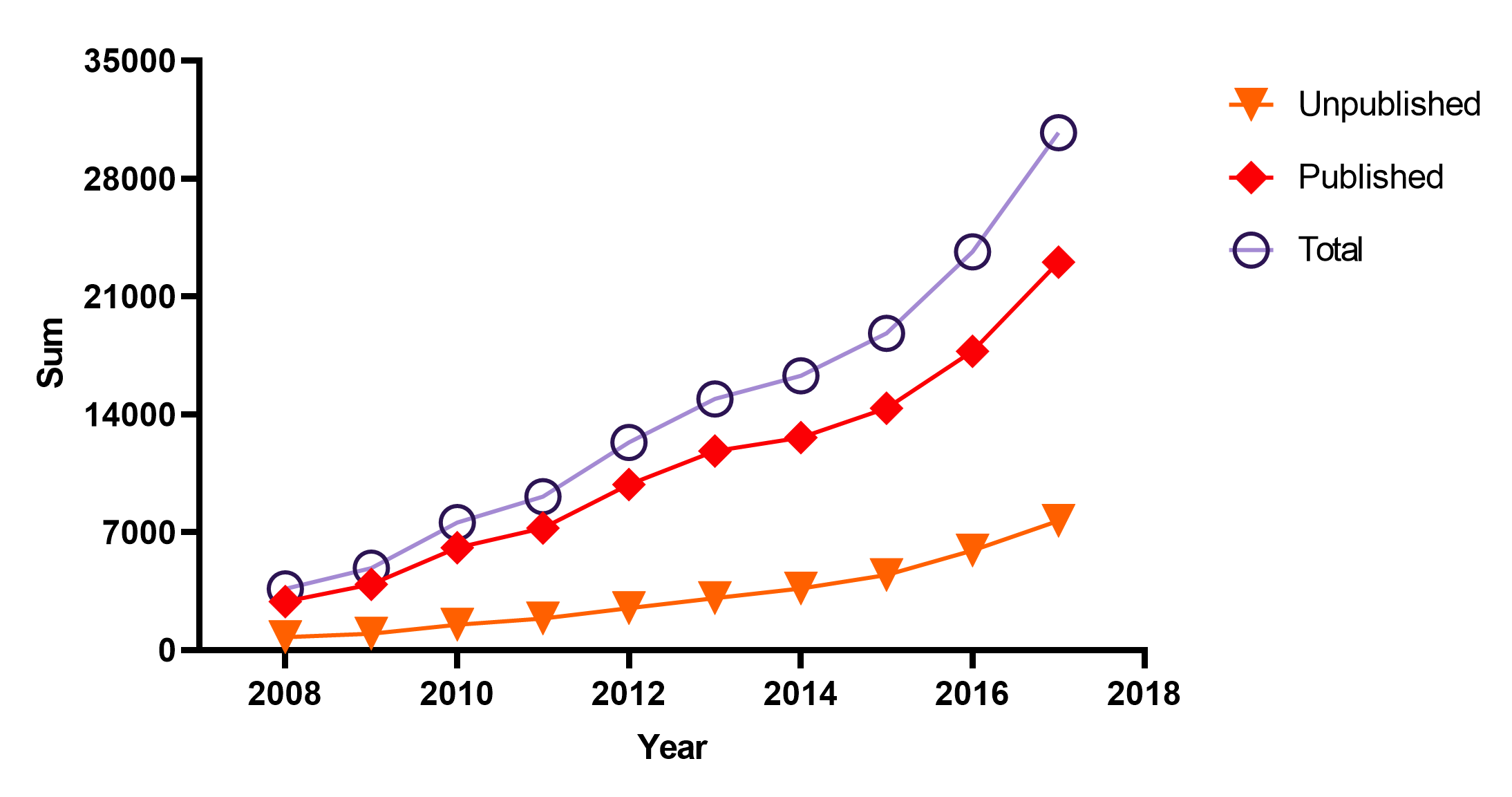}
	\caption{The rise of total, published and unpublished preprints, 2008--2017}
	\label{fig:stat-by-year}
\end{figure}

\begin{figure}[h]
	\centering
	\includegraphics[scale=0.5]{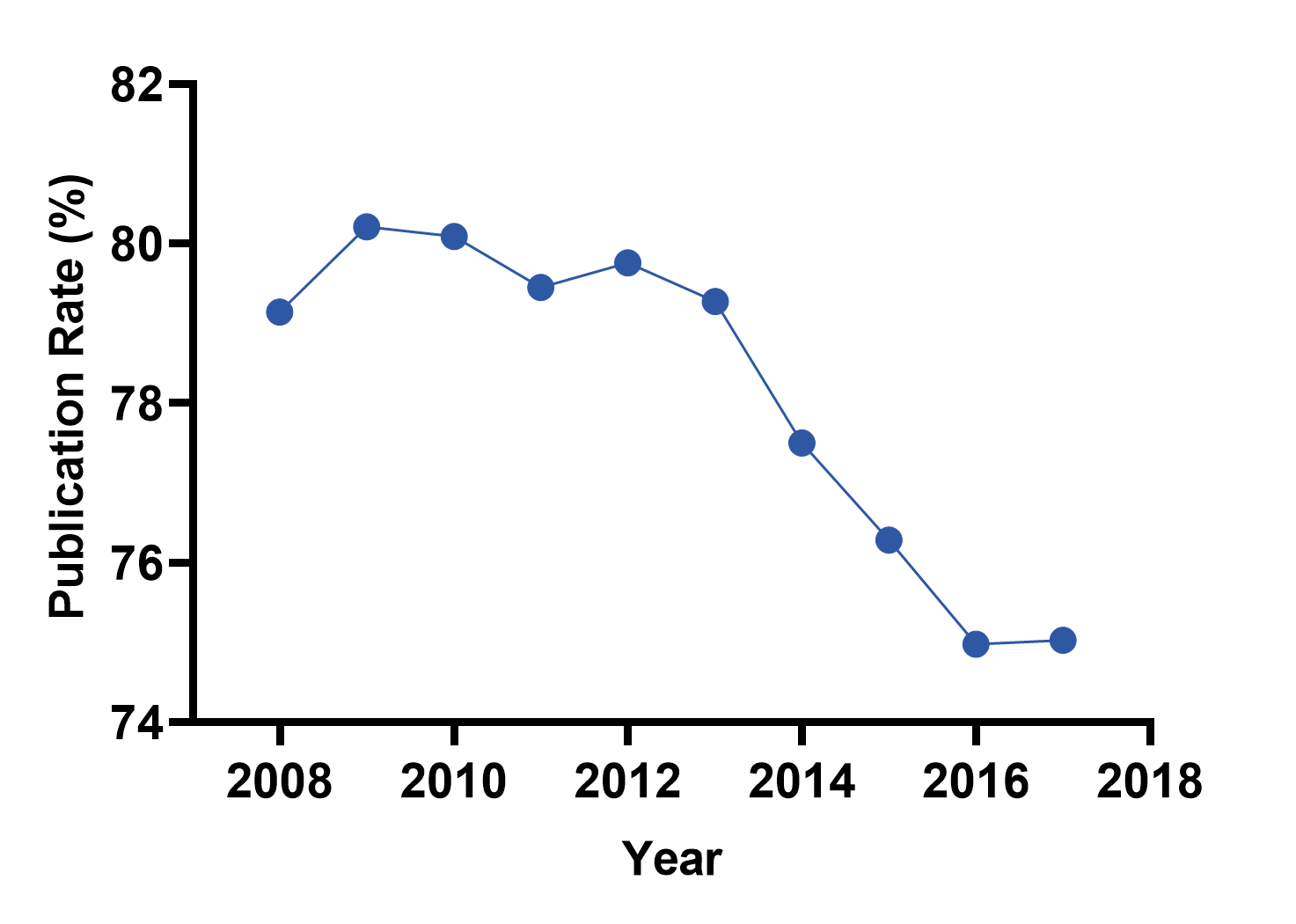}
	\caption{Publication rate of preprints, 2008--2017}
	\label{fig:stat-percen-by-year}
\end{figure}

The statistics of the total, the published, and the unpublished preprints are also presented here by years (see Fig.~\ref{fig:stat-by-year}). The total number of all preprints underwent a significant increase in this ten years' period, soaring from below 5000 in 2008 to over 30,000 in 2017, more than five times. The increase of the submissions has picked up speed since 2015. The increase of published preprints follows the same growing trend. The growth of unpublished preprints, in contrast, only progresses mildly. As shown in Fig.~\ref{fig:stat-percen-by-year}, although the number of preprints has greatly increased in the past decade, the publication rate of preprints has declined as a whole. To some extent, this reveals that preprints have been increasingly popular among researchers. The growth rate of the number of all preprints is much higher than that of finally published ones, that is, the denominator of the publication rate of preprints increases significantly, which eventually leads to a decline in the publication rate of preprints.

\subsection{Subject categories}

\begin{figure}[h]
	\centering
	\includegraphics[scale=0.52]{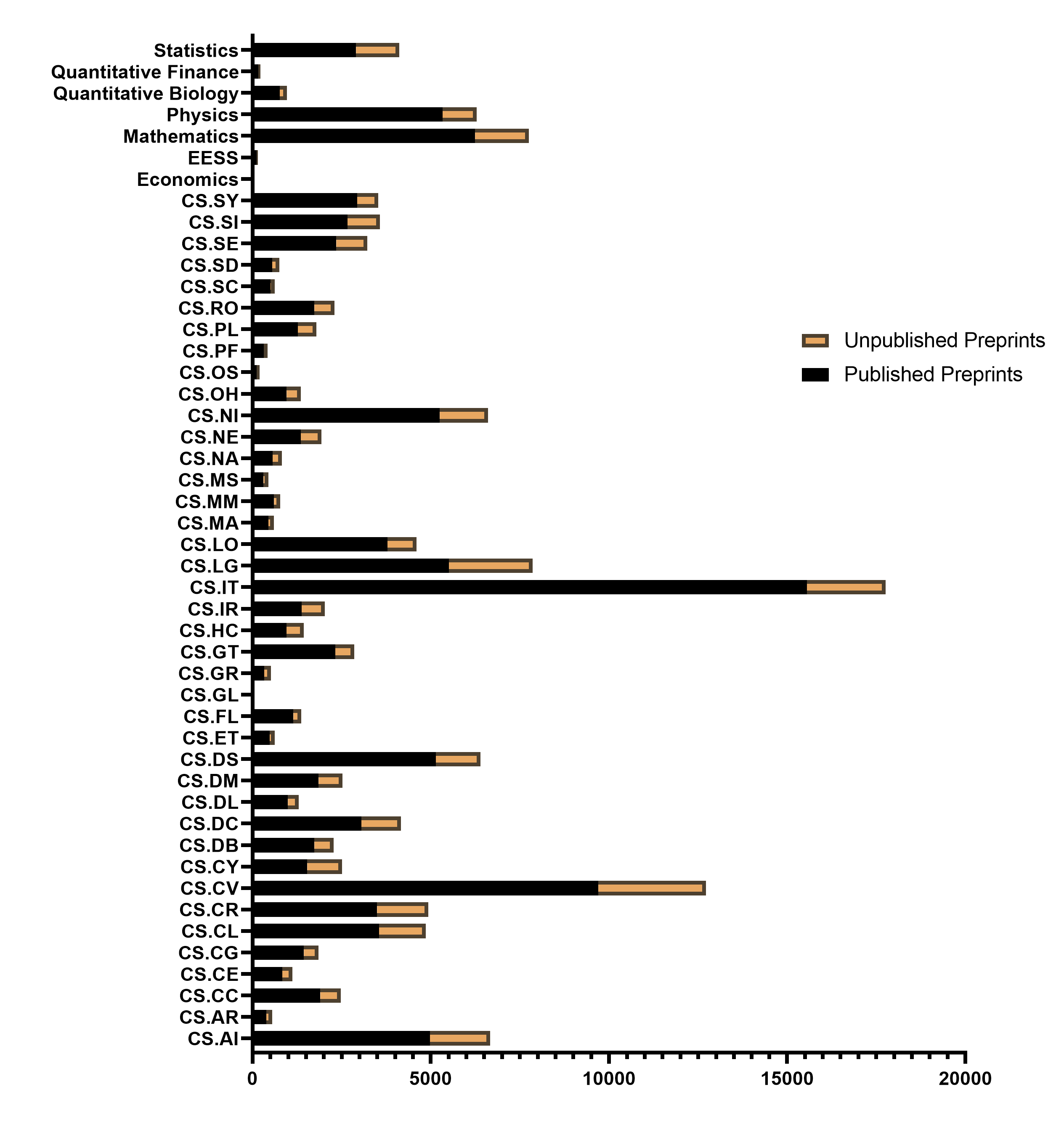}
	\caption{Distribution of published and unpublished preprints by subject category}
	\label{fig:stat-by-subject-category}
\end{figure}

We classified the published and unpublished preprints according to their first category label included in arXiv metadata. For the preprints in the CS field, we divided them into different sub-categories just as arXiv. See Fig.~\ref{fig:stat-by-subject-category} for detailed information.\footnote{Full names of categories on arXiv are attached in Appendix~\ref{sec:app-abbr} for reference.}

From Fig.~\ref{fig:stat-by-subject-category}, we can see that Information Theory (CS.IT) is the most productive category followed by Computer Vision and Pattern Recognition (CS.CV), and Machine Learning (CS.LG). The number of preprints in Information Theory is over twice that in Machine Learning. Altogether, preprints of these top three categories account for about one-fourth of total. Preprints in General Literature (CS.GL) and Operating Systems (CS.OS) account for only a small fraction of the total. Published preprints have a larger proportion than their unpublished counterparts in almost all categories. Mathematics, Physics, and Statistics are the top three categories in terms of the number of CS related preprints. This reveals that cross-disciplined research is prosperous in these three domains.

\subsection{Publication venue}
\begin{figure}[h]
	\centering
	\includegraphics[scale=0.5]{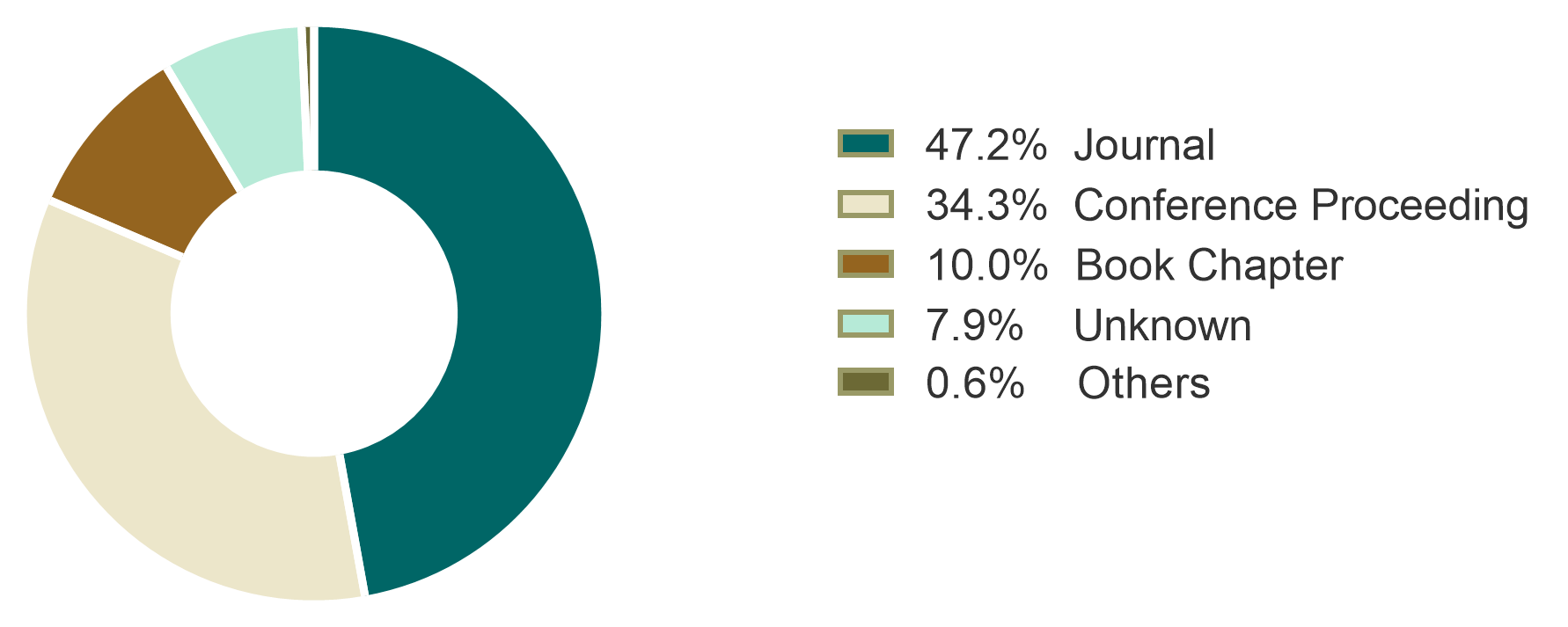}
	\caption{Distribution of preprints by publication venue}
	\label{fig:stat-published-venue}
\end{figure}

As shown in Fig.~\ref{fig:stat-published-venue}, nearly half of the published preprints are journal papers and about one-third are conference papers. Book chapters account for one-tenth of the total. Preprints of unknown publications (information not provided by our data sources) and others only make up a small amount. \citet{vrettas-conferences-2015} found that, on average, top conference papers have a higher citation rate than top journal papers in the domain of CS. However, from the statistical result of our research, more preprints are finally published in journals than in conferences. We suspect the main reason is that most journals have a longer publication period than conference proceedings, so those researchers who want their papers to be published in journals will first submit preprints to arXiv to share their work in advance. On the other hand, owing to the fact that the publication period of conference proceedings is relatively short, there is less need for them to submit preprints.

\subsection{Submission stage}
Submission stage, i.e., the time a preprint is submitted to arXiv is also an important subject matter. When are those preprints submitted to arXiv? Are they submitted before or after formal publication? Would authors commonly upload the formal published version to arXiv? Or do those authors just consider arXiv a platform for quick dissemination of their work to the public?

\begin{figure*}
	\centering
	\subfigure[Journal papers]{\includegraphics[width=2in]{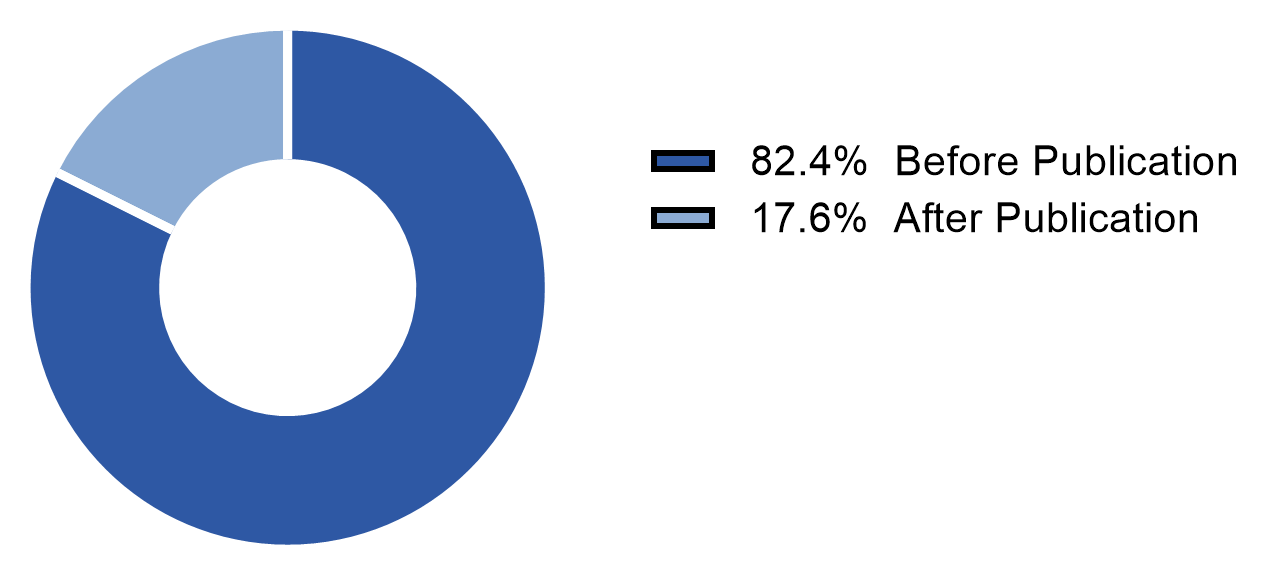}}
	\subfigure[Conference papers]{\includegraphics[width=2in]{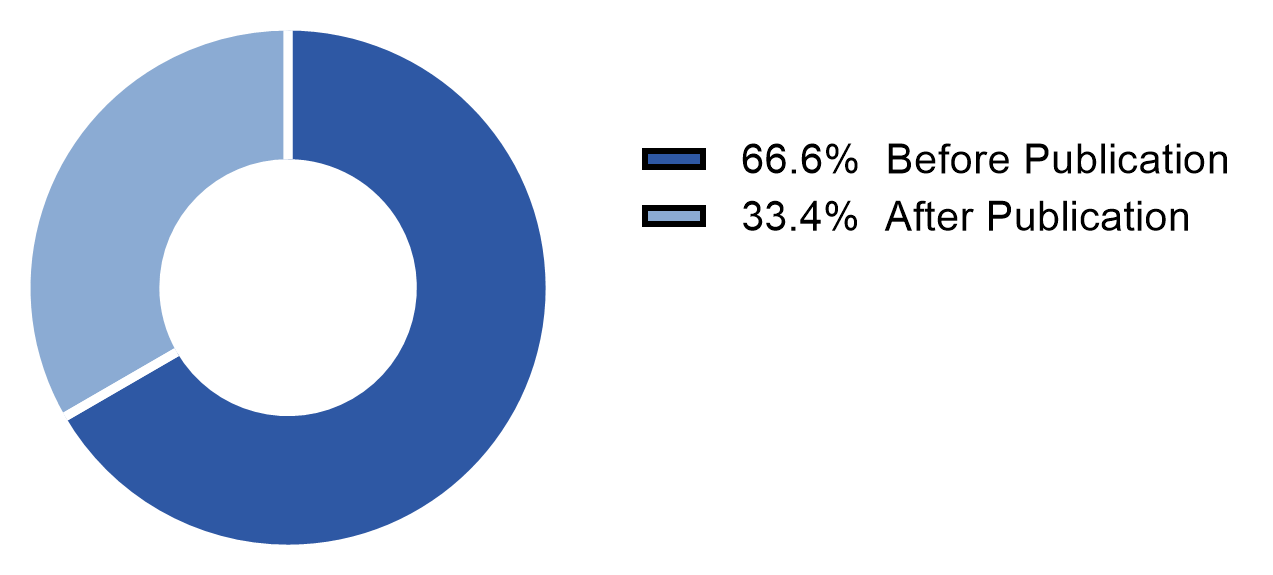}}
	\caption{Distribution of published preprints by submission stage}
	\label{fig:stat-by-submitted-period}
\end{figure*}

We can directly obtain the published date of the peer-reviewed papers from the data sources. However, other information such as received or revised date is contained in the PDF files of the formal published versions. Collecting these data is a difficult task that invites copyright and cost problems. We are finding a solution to get the data and use them to conduct a deeper analysis in the future. At present, we just classified the preprints into two categories according to their created date on arXiv: submitted before publication and submitted after publication. See Fig.~\ref{fig:stat-by-submitted-period} for detailed information.

Fig.~\ref{fig:stat-by-submitted-period} shows that the majority of journal papers and conference papers are submitted before publication. It is reasonable for a preprint server. By comparison, the proportion of journal papers submitted before publication is larger than that of conference papers by 16\%. According to our analysis, this phenomenon exists because it normally takes a longer time for peer review in journal publication than that for conference publication. Therefore, researchers submitting for journal publication are more inclined to post their manuscripts to preprint servers as a claim of priority for the fear of being scooped.

\subsection{Citation count}
Citation count is a vital indicator for the quality of scientific papers. In this section, we compared the citation counts received respectively by the published and unpublished preprints. Citation data were crawled from Google Scholar. If a paper has its arXiv-deposited and published versions indexed separately by Google Scholar, the citation counts of the two versions will be summated. See Table~\ref{table:stat-citation-count} for detailed information. We used the D'Agostino-Pearson test~\citep{dagostino-omnibus-1971,pearson-tests-1977} to test the Normality of citation counts in published preprints, unpublished preprints, journal papers, and conference papers. In the case that the significance level $\alpha$ is predefined as 0.005~\citep{benjamin-redefine-2018}, all $H_0$ are rejected due to smaller probability values (namely \textit{P} value) than $\alpha$, which means these data do not follow Gaussian distribution at the 0.5\% significance level. We then used the Mann-Whitney U test~\citep{mann-test-1947} to compare these data between published preprints and unpublished preprints, journal papers, and conference papers, with the same $\alpha$ as before. The \textit{P} values are all smaller than $\alpha$ indicating that all $H_0$ are rejected, and data in these groups follow different distributions at the 0.5\% significance level. Consequently, the median is chosen as the measure to compare these groups.\footnote{Data in Section~\ref{sec:why-printed} were tested with the same methods and we came to the same conclusion.}

\begin{table}[h]
	\caption{Citations of published and unpublished preprints}
	\label{table:stat-citation-count}
	\centering
	\begin{tabular}{lllll}
	\hline\noalign{\smallskip}
		Items                                     &  Published    &  Journal      &  Conference    &  Unpublished 		\\
	\noalign{\smallskip}\hline\noalign{\smallskip}
		Median citation count                        &  10             &  10               & 10            & 1    \\
		Percentage of zero citations      	         &  11.0\%         &  13.4\%           & 7.6\%         & 37.2\%    \\
	\noalign{\smallskip}\hline
	\end{tabular}
\end{table}

It is obvious from Table~\ref{table:stat-citation-count} that published preprints enjoy higher visibility than unpublished ones. The median citation count of journal papers is the same as conference papers. Over one-third of unpublished preprints have not been cited, while only about one-tenth of published preprints receive zero citations. The percentage of journal papers with zero citations is larger than that of conference papers. There are also unpublished but highly-cited preprints on arXiv. For example, ``ADADELTA: An Adaptive Learning Rate Method''~\citep{zeiler-adadelta-2012}, which introduces an effective gradient descent method, has gained over 3,000 citations.

\section{What preprints can be printed}
\label{sec:why-printed}

In Section~\ref{sec:statistics-analysis}, we analyzed the publication conditions of preprints on arXiv. In this section, comparisons of version history, number of authors and article length, number of references and their citations, number of figures and tables, and proportion of open source code were conducted between the published and unpublished preprints to identify what features enable a preprint to be printed eventually. Based on this comparison and analysis, we went further to provide practical suggestions for academic writers in CS.

Science-parse\footnote{\url{https://github.com/allenai/science-parse}.} is used to parse PDF files on arXiv. The PDF files are transformed into structured XML files with title, authors, abstract, introduction, conclusion, and references included. In order to conduct an in-depth comparison, the published preprints in some subsections were then classified into two categories: conference papers and journal papers. The comparison was performed among published preprints, journal papers, conference papers, and unpublished preprints. In addition, book chapters and other types of publications were excluded from these comparisons. For one thing, book chapters are subject to a writing style greatly different from that of journal and conference papers; for another, papers of other types only account for a tiny share of the total and are thus less representative. For papers published under different titles, their versions deposited on arXiv might not be the final ones, and thus they were also excluded from the data. Apart from that, these comparisons also excluded papers with no updated version submitted to arXiv after the publication so as to ensure the comparison was conducted just among the formal published versions of published preprints.

\subsection{Version history}
ArXiv allows users to make modifications to preprints' content and metadata with no restriction on time. This freedom is one distinct advantage offered by preprint servers, and authors can update their work without going through a complicated review process. We compared the numbers of updates between the published and unpublished preprints.

\begin{table}[h]
	\caption{Proportion of preprints with different version numbers}
	\label{table:stat-modification-num}
	\centering
	\begin{tabular}{lll}
		\hline\noalign{\smallskip}
		Version number               &  Published (\%)        &  Unpublished (\%)          \\
		\noalign{\smallskip}\hline\noalign{\smallskip}
		1 (no update)                &  60.5              &  73.0                    \\
		2                            &  22.9              &  17.7                    \\
		3--5                         &  15.5              &  8.3                     \\
		more than 5                  &  1.1               &  1.0                     \\
		\noalign{\smallskip}\hline
	\end{tabular}
	
\end{table}

Table~\ref{table:stat-modification-num} shows that preprints with one version take up the largest proportion of both published and unpublished preprints. To some extent, this indicates that arXiv is mainly used by researchers as a platform to share their work with others. Published preprints have a lower share of unmodified versions than unpublished preprints, while for the proportion of updating more than one version, published preprints exceed unpublished preprints. This result can be explained by two reasons: (1) repeated revisions normally lead to high quality and thus repeatedly revised preprints have greater chances to be accepted; (2) after their preprints being accepted for publication, most authors will upload the accepted version to arXiv to ensure completeness and consistency of their work. Besides, few preprints on arXiv have more than five versions, and this is because revisions after version five will not be listed in daily mailing anymore.\footnote{\url{https://arxiv.org/help/replace}.}

\subsection{Number of authors and article length}
Number of authors and article length have a huge influence on the first impression of a paper, therefore we conducted comparisons on these two factors. In the comparisons, preprints without certain sections were excluded. See Table~\ref{table:stat-length} for detailed information.

\begin{table}[h]
	\caption{Medians of number of authors and word counts}
	\label{table:stat-length}
	\centering
	\begin{tabular}{lllll}
	\hline\noalign{\smallskip}	
		Items                                  &  Published    &  Journal    &  Conference    &  Unpublished  \\
	\noalign{\smallskip}\hline\noalign{\smallskip}
		Number of authors                      &  3                &  3         &  3             &  2  \\
		Title word count                       &  9                &  9         &  8             &  8  \\
		Abstract word count                    &  150              &  153       &  147           &  138  \\
		Introduction word count                &  691              &  732       &  635           &  563  \\
		Conclusion word count                  &  208              &  245       &  168           &  180  \\
		Acknowledgment word count              &  40               &  44        &  34            &  36  \\
	\noalign{\smallskip}\hline
	\end{tabular}
	
\end{table}

From Table~\ref{table:stat-length}, we can see that the median of the published preprints is higher than that of the unpublished ones in terms of number of authors. This means that multi-authorship is a feature of accepted papers. For article length, the published preprints have all the median values larger than those of unpublished ones. These results illustrate that article length is a quality indicator for reviewers. In particular, the published preprints have significantly longer abstract and introduction, with 9\% and 23\% more in length respectively than those of the unpublished preprints. This demonstrates that detailed abstract and introduction are marked features of published preprints. For the comparison between journal and conference papers, journal papers outnumber conference papers in all items except the number of authors. According to our analysis, the reason for this result is that conference papers have a more rigorous restriction on article length (mostly 8 or 12 pages), thus they are usually in a more concise style.

\subsection{Number of references and their citation counts}
For scientific papers, references are indispensable. To some extent, referencing behaviors are highly correlated to the academic quality of the papers. For this reason, we conducted a comparison of the number of references as well as citation counts received by these references. In order to accomplish the comparison on such a vast quantity of citation counts in a practical way, we only targeted a subset of preprints labeled with Artificial Intelligence from 2016 to 2017. A total of 4,743 preprints were identified within this subset. See Table~\ref{table:stat-reference} for detailed information. Please note that official reference data are not included in the APIs of arXiv. Number of references and their citation counts might thus be a little lower than the actual values due to possibly erroneous parsing of PDF files.

\begin{table}[h]
	\caption{Medians of number of references and citation counts of references}
	\label{table:stat-reference}
	\centering
	\begin{tabular}{lllll}
	\hline\noalign{\smallskip}
		Items                                       &  Published          &  Journal         &  Conference        &  Unpublished  \\
    \noalign{\smallskip}\hline\noalign{\smallskip}
		Number of references                        &  30                   &  35                 &  29              &  23  \\
		Citation counts of references               &  34,753               &  25,683             &  42,273          &  23,905  \\
    \noalign{\smallskip}\hline
    \end{tabular}
	
\end{table}

It is shown clearly in Table~\ref{table:stat-reference} that compared with the unpublished preprints, the published preprints have more references. This result indicates that the number of references is positively related to the acceptance of papers. In terms of the median number of references, the published preprints cite 30\% more than that of the unpublished ones. The median citation counts of the published preprints' references is also 45\% higher than that of the unpublished ones. Judging from the median, the journal papers have more references than the conference papers, while the conference papers have more highly cited references.

Citation counts of references are rather high, and this is because they are pushed up by some most cited references. For example, ``R: A Language and Environment for Statistical Computing''~\citep{rct-r-2011} has received more than 140,000 citations.

\subsection{Number of figures and tables}
Figures and tables are two essential components in academic writing. They can highlight and reinforce the key information in a straightforward way so that the paper can be more reader-friendly. Figures and tables were parsed, counted in number, and calculated for their median values separately. See Table~\ref{table:stat-materials} for detailed information.

\begin{table}[h]
	\caption{Medians of number of figures and number of tables}
	\label{table:stat-materials}
	\centering
	\begin{tabular}{lllll}
		\hline\noalign{\smallskip}
		Items                         &  Published         &  Journal        &  Conference         &  Unpublished  \\
		\noalign{\smallskip}\hline\noalign{\smallskip}
		Number of figures             &  4                    &  5                  &  4                 &  4  \\
		Number of tables              &  1                    &  0                  &  1                 &  0  \\
		\noalign{\smallskip}\hline
	\end{tabular}

\end{table}

The results shown in Table~\ref{table:stat-materials} are different from what we expected. The published and unpublished preprints score the same in the median number of figures. The journal papers and the unpublished preprints both surprisingly have zero as the median number of tables. We were afraid that these values were caused by error automatically parsing steps, so we manually calculated the number of tables in PDF files for 100 randomly selected samples from the unpublished preprints and the result remains as zero.\footnote{There were 53 preprints without tables among these 100 records, and we indeed found that a small part of the automatically parsing values was smaller than the manually calculated ones.} It is also worth noting that journal papers use more figures and fewer tables than conference papers. Overall, the papers published do not necessarily feature a larger number in figures and tables. However, we can reach a conclusion from these results that CS papers as a whole normally feature the use of figures, which shows that researchers nowadays are well aware of the effectiveness of figures as a form of illustration.

\subsection{Open source code}
The reproducibility of CS research is largely based on the availability of its source code, and thus whether the source code is provided can be considered an indicator for the reliability and credibility of the research. Opening source code can be a solid proof for the confidence of researchers to their academic work as others can thus reproduce the results. In this section, statistical analysis was performed to determine whether opening source would influence the acceptance rate. We counted the respective percentage of open source papers for the published and unpublished preprints.

We conducted a mapping between the sampled preprints and their corresponding code repositories using Papers With Code. Altogether 5,319 preprints were identified with open source code provided, which only accounted for 3.7\% of the total sample preprints. One explanation is that papers in some domains of CS are purely theoretical and thus involve no code. Therefore, we only took into consideration those preprints labeled with at least one of the following categories: Artificial Intelligence, Computation and Language, Computer Vision and Pattern Recognition, Information Retrieval, Machine Learning, and Neural and Evolutionary Computing. A total of 46,937 preprints were identified, among which, the percentage was 11.3\%. The percentage was still relatively low. An explanation is that Papers With Code prefers to index up-to-date research, and thus some of our sampled preprints selected from 2008 to 2017 might not be covered by Papers With Code.

Among the preprints with open source code, 79.7\% have been accepted by peer-reviewed venues. It is strong evidence that opening source code correlates tightly with the acceptance rate. Therefore, we suggest researchers provide open source code in their papers.

\section{Future work}

For future study, we are looking forward to continuing our work in the following directions. First and foremost, we hope to extend our research from CS to other domains and from arXiv to other preprint servers. Next, we are exploring a more efficient solution to conduct quantitative analysis on citations. The current method we adopted is relatively time-consuming and costly, thus we only analyzed Artificial Intelligence preprints in 2016--2017. With a new solution, we can extend our research to cover preprints in more fields and a longer time range. Last but not least, we would love to include other factors, chiefly the influence of funding, structure, and even content in our comparison between published preprints and unpublished ones.

\section{Conclusion}

In this paper, we introduce a deep learning-based method to map arXiv-deposited preprints to their corresponding published versions with different titles in peer-reviewed venues. With the help of this enabling method and our data sources, we found that 66\% of CS preprints submitted to arXiv between 2008 and 2017 have been published with the same title, and 11\% are published under different titles and with other modifications. These results show that posting manuscripts to preprint servers contributes to the acceptance of papers in peer-reviewed venues. Among these published preprints, nearly half of them are published in journals, and around one-third of them are accepted in conference proceedings. Apart from that, we went further to analyze the differences between published and unpublished preprints. The results demonstrate that, compared with unpublished preprints, most of the published preprints in the CS domain share common features like adequate revisions, multiple authorship, detailed abstract and introduction, extensive and authoritative references, and available source code.

\section*{Acknowledgments}

We are grateful to Yingmin Wang and Xingchen He for their assistance in processing data for this research. We appreciated Ziyi Chen for her help of statistical analysis. We also thank two anonymous reviewers for their insightful comments. Special and heartfelt gratitude goes to the first author's wife Fenmei Zhou, for her understanding and love. Her unwavering support and continuous encouragement enable this research to be possible.

\section*{Funding}

This work is partly sponsored by the State Language Commission of China through the 13th Five-Year Plan project Artificial Intelligence and Language (Grant No. WT135-38).

\section*{Compliance with ethical standards}

\textbf{Conflict of interest} The authors declare that they have no conflict of interest.

\bigskip
\bigskip

\bibliography{sn-article}    

\begin{thebibliography}{30}
\providecommand{\natexlab}[1]{#1}
\providecommand{\url}[1]{{#1}}
\providecommand{\urlprefix}{URL }
\providecommand{\doi}[1]{\url{https://doi.org/#1}}
\providecommand{\eprint}[2][]{\url{#2}}
 \bibcommenthead

\bibitem[{Annesley et~al.(2017)Annesley, Scott, Bastian, Fonseca, Ioannidis,
  Keller, and Polka}]{annesley-biomedical-2017}
Annesley T, Scott M, Bastian H, et~al. (2017) {Biomedical journals and preprint
  services: Friends or foes?} Clinical Chemistry 63(2):453--458.
  \doi{10.1373/clinchem.2016.268227}

\bibitem[{Beltagy et~al.(2019)Beltagy, Lo, and Cohan}]{beltagy-scibert-2019}
Beltagy I, Lo K, Cohan A (2019) {SciBERT: A pretrained language model for
  scientific text}. In: EMNLP-IJCNLP, \doi{10.18653/v1/D19-1371}

\bibitem[{Benjamin et~al.(2018)Benjamin, Berger, Johannesson, Nosek,
  Wagenmakers, Berk, Bollen, Brembs, Brown, Camerer, Cesarini, Chambers, Clyde,
  Cook, De~Boeck, Dienes, Dreber, Easwaran, Efferson, Fehr, Fidler, Field,
  Forster, George, Gonzalez, Goodman, Green, Green, Greenwald, Hadfield,
  Hedges, Held, Hua~Ho, Hoijtink, Hruschka, Imai, Imbens, Ioannidis, Jeon,
  Jones, Kirchler, Laibson, List, Little, Lupia, Machery, Maxwell, McCarthy,
  Moore, Morgan, Munafó, Nakagawa, Nyhan, Parker, Pericchi, Perugini, Rouder,
  Rousseau, Savalei, Schönbrodt, Sellke, Sinclair, Tingley, Van~Zandt, Vazire,
  Watts, Winship, Wolpert, Xie, Young, Zinman, and
  Johnson}]{benjamin-redefine-2018}
Benjamin DJ, Berger JO, Johannesson M, et~al. (2018) {Redefine statistical
  significance}. Nature Human Behaviour 2(1):6--10.
  \doi{10.1038/s41562-017-0189-z}

\bibitem[{Berg et~al.(2016)Berg, Bhalla, Bourne, Chalfie, Drubin, Fraser,
  Greider, Hendricks, Jones, Kiley, King, Kirschner, Krumholz, Lehmann, Leptin,
  Pulverer, Rosenzweig, Spiro, Stebbins, Strasser, Swaminathan, Turner, Vale,
  VijayRaghavan, and Wolberger}]{berg-preprints-2016}
Berg JM, Bhalla N, Bourne PE, et~al. (2016) {Preprints for the life sciences}.
  Science 352(6288):899--901. \doi{10.1126/science.aaf9133}

\bibitem[{Björk and Solomon(2013)}]{bjork-publishing-2013}
Björk BC, Solomon D (2013) {The publishing delay in scholarly peer-reviewed
  journals}. Journal of Informetrics 7(4):914--923.
  \doi{10.1016/j.joi.2013.09.001}

\bibitem[{Brody et~al.(2006)Brody, Harnad, and Carr}]{brody-earlier-2006}
Brody T, Harnad S, Carr L (2006) {Earlier web usage statistics as predictors of
  later citation impact}. Journal of the American Society for Information
  Science and Technology 57(8):1060--1072. \doi{10.1002/asi.20373}

\bibitem[{Cavacini(2015)}]{cavacini-what-2015}
Cavacini A (2015) {What is the best database for computer science journal
  articles?} Scientometrics 102(3):2059--2071. \doi{10.1007/s11192-014-1506-1}

\bibitem[{D'Agostino(1971)}]{dagostino-omnibus-1971}
D'Agostino RB (1971) {An omnibus test of normality for moderate and large size
  samples}. Biometrika 58(2):341--348. \doi{10.1093/biomet/58.2.341}

\bibitem[{Davis and Fromerth(2007)}]{davis-does-2007}
Davis PM, Fromerth MJ (2007) {Does the arXiv lead to higher citations and
  reduced publisher downloads for mathematics articles?} Scientometrics
  71(2):203--215. \doi{10.1007/s11192-007-1661-8}

\bibitem[{Devlin et~al.(2019)Devlin, Chang, Lee, and
  Toutanova}]{devlin-bert-2019}
Devlin J, Chang MW, Lee K, et~al. (2019) {BERT: Pre-training of deep
  bidirectional transformers for language understanding}. In: NAACL-HLT,
  \doi{10.18653/v1/N19-1423}

\bibitem[{Feldman et~al.(2018)Feldman, Lo, and Ammar}]{feldman-citation-2018}
Feldman S, Lo K, Ammar W (2018) {Citation count analysis for papers with
  preprints}. arXiv preprint arXiv:1805.05238

\bibitem[{Ginsparg(2011)}]{ginsparg-arxiv-2011}
Ginsparg P (2011) {ArXiv at 20}. Nature 476(7359):145--147.
  \doi{10.1038/476145a}

\bibitem[{Halpern(2000)}]{halpern-corr-2000}
Halpern JY (2000) {CoRR: A computing research repository}. ACM Journal of
  Computer Documentation 24(2):41--48. \doi{10.1145/337271.337274}

\bibitem[{Kling(2004)}]{kling-internet-2004}
Kling R (2004) {The internet and unrefereed scholarly publishing}. Annual
  Review of Information Science and Technology 38(1):591--631.
  \doi{10.1002/aris.1440380113}

\bibitem[{Kousha and Thelwall(2008)}]{kousha-sources-2008}
Kousha K, Thelwall M (2008) {Sources of Google Scholar citations outside the
  Science Citation Index: A comparison between four science disciplines}.
  Scientometrics 74(2):273--294. \doi{10.1007/s11192-008-0217-x}

\bibitem[{Lammey(2014)}]{lammey-crossref-2014}
Lammey R (2014) {CrossRef developments and initiatives: An update on services
  for the scholarly publishing community from CrossRef}. Science Editing
  1(1):13--18. \doi{10.6087/kcse.2014.1.13}

\bibitem[{Larivière et~al.(2014)Larivière, Sugimoto, Macaluso, Milojević,
  Cronin, and Thelwall}]{lariviere-arxiv-2014}
Larivière V, Sugimoto CR, Macaluso B, et~al. (2014) {arXiv e-prints and the
  journal of record: An analysis of roles and relationships}. Journal of the
  Association for Information Science and Technology 65(6):1157--1169.
  \doi{10.1002/asi.23044}

\bibitem[{Ley(2002)}]{ley-dblp-2002}
Ley M (2002) {The DBLP computer science bibliography: Evolution, research
  issues, perspectives}. In: SPIRE, \doi{10.1007/3-540-45735-6_1}

\bibitem[{Mann and Whitney(1947)}]{mann-test-1947}
Mann HB, Whitney DR (1947) {On a test of whether one of two random variables is
  stochastically larger than the other}. The Annals of Mathematical Statistics
  18(1):50--60. \doi{10.1214/aoms/1177730491}

\bibitem[{Martin-Martin et~al.(2017)Martin-Martin, Orduna-Malea, Harzing, and
  Delgado López-Cózar}]{martin-can-2017}
Martin-Martin A, Orduna-Malea E, Harzing AW, et~al. (2017) {Can we use Google
  Scholar to identify highly-cited documents?} Journal of Informetrics
  11(1):152--163. \doi{10.1016/j.joi.2016.11.008}

\bibitem[{Martín-Martín et~al.(2018)Martín-Martín, Orduna-Malea, Thelwall,
  and Delgado López-Cózar}]{martin-google-2018}
Martín-Martín A, Orduna-Malea E, Thelwall M, et~al. (2018) {Google Scholar,
  Web of Science, and Scopus: A systematic comparison of citations in 252
  subject categories}. Journal of Informetrics 12(4):1160--1177.
  \doi{10.1016/j.joi.2018.09.002}

\bibitem[{Pearson et~al.(1977)Pearson, D'Agostino, and
  Bowman}]{pearson-tests-1977}
Pearson ES, D'Agostino RB, Bowman KO (1977) {Tests for departure from
  normality: Comparison of powers}. Biometrika 64(2):231--246.
  \doi{10.1093/biomet/64.2.231}

\bibitem[{Rieger et~al.(2016)Rieger, Steinhart, and Cooper}]{rieger-arxiv-2016}
Rieger OY, Steinhart G, Cooper D (2016) {arXiv@25: Key findings of a user
  survey}. arXiv preprint arXiv:1607.08212

\bibitem[{Sutton and Gong(2017)}]{sutton-popularity-2017}
Sutton C, Gong L (2017) {Popularity of arXiv.org within computer science}.
  arXiv preprint arXiv:1710.05225

\bibitem[{Team(2013)}]{rct-r-2011}
Team RC (2013) {R: A language and environment for statistical computing}. R
  Foundation for Statistical Computing

\bibitem[{Vale(2015)}]{vale-accelerating-2015}
Vale RD (2015) {Accelerating scientific publication in biology}. Proceedings of
  the National Academy of Sciences of the United States of America
  112(44):13,439--13,446. \doi{10.1073/pnas.1511912112}

\bibitem[{Van~Noorden(2014)}]{van-arxiv-2014}
Van~Noorden R (2014) {The arXiv preprint server hits 1 million articles}.
  Nature \doi{10.1038/nature.2014.16643}

\bibitem[{Vrettas and Sanderson(2015)}]{vrettas-conferences-2015}
Vrettas G, Sanderson M (2015) {Conferences versus journals in computer
  science}. Journal of the Association for Information Science and Technology
  66(12):2674--2684. \doi{10.1002/asi.23349}

\bibitem[{Warner(2001)}]{warner-open-2001}
Warner S (2001) {Open Archives Initiative protocol development and
  implementation at arXiv}. arXiv preprint arXiv:cs/0101027

\bibitem[{Zeiler(2012)}]{zeiler-adadelta-2012}
Zeiler MD (2012) {ADADELTA: An adaptive learning rate method}. arXiv preprint
  arXiv:1212.5701

\end{thebibliography}


\begin{appendices}

\section{Abbreviation---full name of arXiv categories}
\label{sec:app-abbr}

\begin{table}[h]
	\centering 
	\resizebox{0.98\textwidth}{!}{
	\begin{tabular}{|l|l|l|l|}
		\hline
	Abbr. 		&  Full name 		&  Abbr.  		&  Full name  \\
	\hline
	CS.AI	&	Artificial Intelligence 	&	CS.IR	&	Information Retrieval			\\
	CS.AR	&	Hardware Architecture		&	CS.IT	&	Information Theory			\\
	CS.CC	&	Computational Complexity	&	CS.LG	&	Machine Learning 			\\
	CS.CE	&	Computational Engineering, Finance, and Science	&	CS.LO	&	Logic in Computer Science			\\
	CS.CG	&	Computational Geometry		&	CS.MA	&	Multiagent Systems			\\
	CS.CL	&	Computation and Language 	&	CS.MM	&	Multimedia			\\
	CS.CR	&	Cryptography and Security	&	CS.MS	&	Mathematical Software			\\
	CS.CV	&	Computer Vision and Pattern Recognition	&	CS.NA	&	Numerical Analysis			\\
	CS.CY	&	Computers and Society		&	CS.NE	&	Neural and Evolutionary Computation			\\
	CS.DB	&	Databases					&	CS.NI	&	Networking and Internet Architecture			\\
	CS.DC	&	Distributed, Parallel, and Cluster Computing	&	CS.OH	&	Other			\\
	CS.DL	&	Digital Libraries			&	CS.OS	&	Operating Systems			\\
	CS.DM	&	Discrete Mathematics		&	CS.PF	&	Performance			\\
	CS.DS	&	Data Structures and Algorithms	&	CS.PL	&	Programming Languages			\\
	CS.ET	&	Emerging Technologies 		&	CS.RO	&	Robotics			\\
	CS.FL	&	Formal Languages and Automata Theory 	&	CS.SC	&	Symbolic Computation			\\
	CS.GL	&	General Literature 			&	CS.SD	&	Sound			\\
	CS.GR	&	Graphics					&	CS.SE	&	Software Engineering			\\
	CS.GT	&	Computer Science and Game Theory	&	CS.SI	&	Social and Information Networks			\\
	CS.HC	&	Human-Computer Interaction	&	CS.SY	&	Systems and Control			\\
	EESS	&	Electrical Engineering and Systems Science	&	        	&			\\
	\hline
		
	\end{tabular}
	}
\end{table}

\end{appendices}

\end{document}